\documentclass[preprint2]{aastex}

\usepackage{amsmath}
\usepackage{graphicx}
\usepackage{epsfig,psfrag,epic,eepic}
\usepackage{textcomp}
\usepackage{times}
\usepackage{longtable}

\slugcomment{Accepted for publication in the Astrophysical Journal Letters}

\shorttitle{Brown dwarf companions in Sco-Cen}
\shortauthors{Janson et al.}

\begin{document}

\title{New Brown Dwarf Companions to Young Stars in Scorpius-Centaurus\altaffilmark{*}}

\author{Markus Janson\altaffilmark{1,7}, 
Ray Jayawardhana\altaffilmark{2}, 
Julien H. Girard\altaffilmark{3}, 
David Lafreni{\`e}re\altaffilmark{4}, 
Mariangela Bonavita\altaffilmark{2}, 
John Gizis\altaffilmark{5},
Alexis Brandeker\altaffilmark{6}
}

\altaffiltext{*}{Based on Gemini observations from programs GS-2011A-Q-44, GS-2012A-Q-18, and GS-2012A-DD-6, and on ESO observations from program 089.C-0422(A).}
\altaffiltext{1}{Princeton University, Princeton, NJ, USA;\\ \texttt{janson@astro.princeton.edu}}
\altaffiltext{2}{University of Toronto, Toronto, ON, Canada}
\altaffiltext{3}{European Southern Observatory, Santiago, Chile}
\altaffiltext{4}{University of Montreal, Montreal, QC, Canada}
\altaffiltext{5}{University of Delaware, Newark, DE, USA}
\altaffiltext{6}{Stockholm University, Stockholm, Sweden}
\altaffiltext{7}{Hubble fellow}

\begin{abstract}\noindent
We present the discoveries of three faint companions to young stars in the Scorpius-Centaurus region, imaged with the NICI instrument on Gemini South. We have confirmed all three companions through common proper motion tests. Follow-up spectroscopy has confirmed two of them, HIP~65423~B and HIP~65517~B, to be brown dwarfs, while the third, HIP~72099~B, is more likely a very low-mass star just above the hydrogen burning limit. The detection of wide companions in the mass range of $\sim$40--100~$M_{\rm jup}$ complements previous work in the same region, reporting detections of similarly wide companions with lower masses, in the range of $\sim$10--30$M_{\rm jup}$. Such low masses near the deuterium burning limit have raised the question of whether those objects formed like planets or stars. The existence of intermediate objects as reported here could represent a bridge between lower-mass companions and stellar companions, but in any case demonstrate that mass alone may not provide a clear-cut distinction for the formation of low-mass companions to stars.

\end{abstract}


\section{Introduction}

Although a number of brown dwarfs are known to orbit stars \citep[e.g.][]{nakajima1995,burgasser2000}, most detected brown dwarfs are single \citep[e.g.][]{kirkpatrick2012} or accompanying other brown dwarfs \citep[e.g.][]{bouy2008}. In general, brown dwarf companions to stars are rare, particularly at small orbits of a few AU \citep[e.g.][]{grether2006}. However, indications exist that they may be at least slightly more common in wider configurations \citep[e.g.][]{lafreniere2011,ireland2011}.

The Sco-Cen region \citep{dezeeuw1999} is a young stellar association consisting of the sub-regions Upper Scorpius (USco), Upper Centaurus Lupus (UCL) and Lower Centaurus Crux (LCC). Age estimates of the various sub-regions have varied in the literature \citep{dezeeuw1999,pecaut2012}, but recent direct comparisons to other young groups have shown that the whole region is younger than the $\beta$~Pic moving group (12~Myr), placing the UCL and LCC sub-groups at ages of approximately 10~Myr, and USco probably somewhat younger \citep{song2012}. Previous detections of wide ($\sim$200--600~AU) low-mass substellar companions in USco \citep[e.g.][]{lafreniere2008,ireland2011} in shallow adaptive optics (AO) surveys motivated us to perform an extended survey in the whole Sco-Cen region using NICI at Gemini South. In the course of this study, we detected several new companion candidates that are being followed up for astrometric and spectroscopic confirmation. In this Letter, we report the discovery of three brown dwarf candidates, all of which have been confirmed though common proper motion and two of which have also been spectroscopically confirmed to be substellar. Two of the host stars, HIP~65423 and HIP~65517, are members of LCC, and HIP~72099 is a member of UCL \citep{dezeeuw1999,chen2011}. HIP~72099 has a known debris disk \citep{chen2011}.

\section{Observations and Data Reduction}

\subsection{Imaging Observations}

Images were taken with NICI at Gemini South as part of a program to search for planet and brown dwarf companions to stars in the Sco-Cen region. The dual band imaging mode was used, with the broad $K_s$ filter (2.15~$\mu$m with 14.5\% bandwidth) in the red channel and the narrow $H_2$ filter (2.12~$\mu$m with 1.23\% bandwidth) in the blue channel. Each target was observed with a short exposure of 0.38~s integration time and 10 coadds, and a longer exposure of 80~s in a single integration. This setup was used to allow for a wide dynamic range with sensitivity to bright and very close-in companions in the short exposure and narrow band, as well as faint and wide companions in the long exposure and broad band. The targets presented here were each first observed in April or June 2011 and then re-observed on both 2012 Feb 29 and 2012 Apr 2.

The data were reduced with a custom IDL pipeline. Flat fielding and dark subtraction was applied to all individual frames. While no dithering was used, observations of different targets in the survey were taken contiguously and on different random places on the detector, thus forming an equivalent of the often used `jittering' strategy where the science frames of other targets were used as references for a given target frame, for sky subtraction and identification of bad pixels. All frames were then distortion corrected and re-oriented to a common orientation with North up and East to the left\footnote{By default, the NICI images are rotated with North pointing downwards, and since the NICI port varied between up-looking and side-looking from 2011 to 2012, the orientations varied between a right-handed and a left-handed configuration.}. Ensemble astrometry showed that a systematic 0.3$^{\rm o}$ offset was present during one specific NICI mounting in June 2011, so since one epoch of imaging of both HIP~65517 and HIP~72099 was taken during that period, those images have been corrected to account for this offset.

\begin{figure}[p]
\centering
\includegraphics[width=6cm]{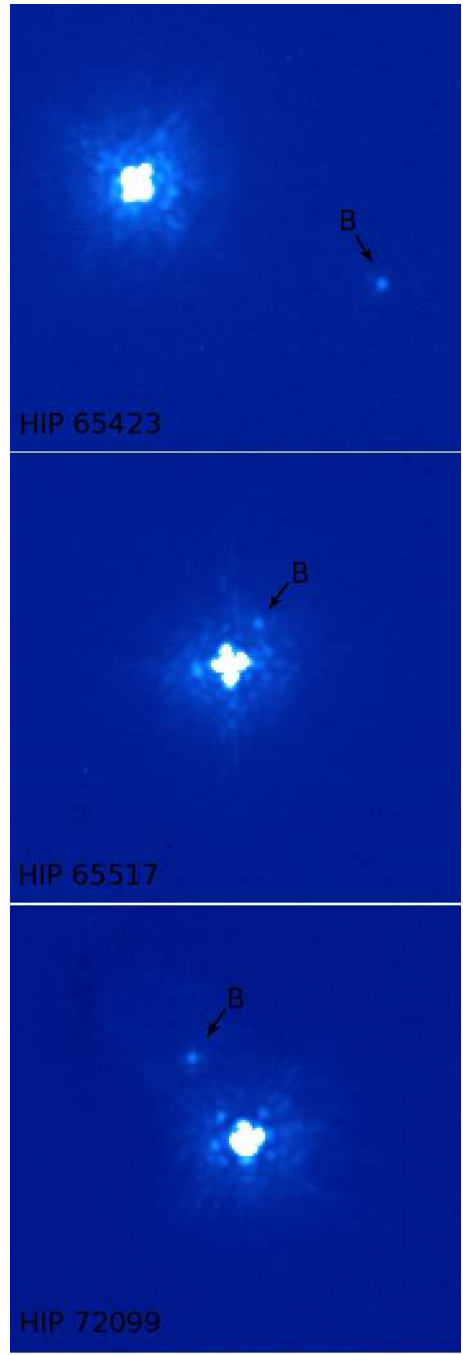}
\caption{Discovery images of the three systems. The projected separations range from 0.350\arcsec for HIP~65517~B to 1.835\arcsec for HIP~65423~B. North is up and East is to the left. Note that apparent point sources closer to the stars are due to known ghosts.}
\label{f:discim}
\end{figure}

Reduced images for the three systems are shown in Fig. \ref{f:discim}. The limiting magnitudes range from 19.4 to 20.2~mag  in the 80~s broad-band images, and no tertiary candidates were found in either system.

\subsection{Spectroscopic Observations}

Spectra for the three targets were acquired using NACO at the VLT in May 2012 for HIP~65423 and HIP~65517 and July 2012 for HIP~72099. We used a slit of 172~mas width and 40\arcsec\ length in the $H+K$ setting giving spectra over a wavelength range of $\sim$1.33--2.53$\mu$m simultaneously at a spectral resolution of 550. Both the primary and secondary were included in the slit for all exposures. The observations were executed using an ABBA nodding pattern with a nod throw of 10\arcsec. One AB pair was taken with a short integration time of 1~s with 12 coadds for getting the primary star in a fully linear count regime, followed by two full ABBA cycles with longer integrations of 5~s with 20 coadds per frame, for a total of 800~s on each target.

Data reduction was performed using a custom IDL pipeline based on a previously developed code for NACO spectroscopy \citep{janson2010}. The data were flat fielded and dark subtracted, and each contiguous AB or BA set was pairwise subtracted to remove the sky. For each individual frame, a trace of the spectrum of the primary star was acquired through gaussian centroiding on each individual pixel row across the wavelength direction. This worked well, even though mild saturation occured in the $H$-band range for the long exposures. The traces were then re-centered using spline interpolation to form a straight vertical trace centered on a common pixel column for all frames, after which the short exposures were coadded to form a single non-saturated primary star reference frame, and likewise the long frames were coadded to form a single frame with maximal sensitivity to the companion. 

Three 1D spectra are extracted from the spectral trace image, each in an aperture of 3 pixels: One centered on the primary (this is done on the basis of the short-exposure coadded frame, since the primary star is partly saturated otherwise), one centered on the companion, and another centered on the exact opposite side of the primary, to represent the PSF at the location of the companion. Wavelength solutions for these spectra are acquired using both sky lines and intrinsic lines in the target spectra. The PSF reference spectrum is then subtracted from the companion spectrum to form a PSF-subtracted companion spectrum. This in turn is divided by the primary spectrum, appropriately scaled to account for the difference in integration time between the short and long exposures, which gives a telluric-free contrast spectrum. By multiplying the contrast spectrum with a model spectrum for the primary, this yields a companion spectrum that is telluric-free and subject only to the uncertainties of the primary star properties. In our case, we simply employ a blackbody spectrum at the stellar temperature as a model for the stellar flux, since the intrinsic stellar lines are weak in this wavelength range and do not affect the purely continuum-based analysis that we perform in Sect. \ref{s:spectra}. The adopted temperatures are 5830~K for HIP~65423 (G3V), 5860~K for HIP~65517 (G2V), and 6360~K for HIP~72099 (F6V), using SIMBAD spectral types.

\section{Analysis}

\subsection{Astrometric Analysis}

The main limiting issue for accurate astrometry for the NICI observations is that the minimum integration time is 0.38~s, which for the bright stars observed in our program means inevitable saturation. Furthermore, the NICI PSF is strongly asymmetric and the non-linearity response of the detector is complex. This means that neither Moffat fitting of the PSF halo nor centroiding on the saturated core, which often work well for similar instruments like HiCIAO \citep[e.g.][]{thalmann2011} or NACO \citep[e.g.][]{janson2008}, can be satisfactorily applied here. Furthermore, the PSF spider arms, which have been successfully used for e.g. NIRI \citep[e.g.][]{lafreniere2007} are not concentric for NICI. Yet, it is fairly straightforward to determine where the actual center is by eye. We have tried all these techniques on $\sim$50 target stars with suspected background stars in the program, and found that manual centering gave the smallest scatter in measured versus expected background motion, meaning it is likely the most accurate method. Hence, we use that for HIP~65423, HIP~65517 and HIP~72099 as well. As we will see below, this indeed leads to a good astrometric quality. For astrometry of the companion candidates, we used Gaussian centroiding.

\begin{figure*}[p]
\centering
\includegraphics[width=10cm]{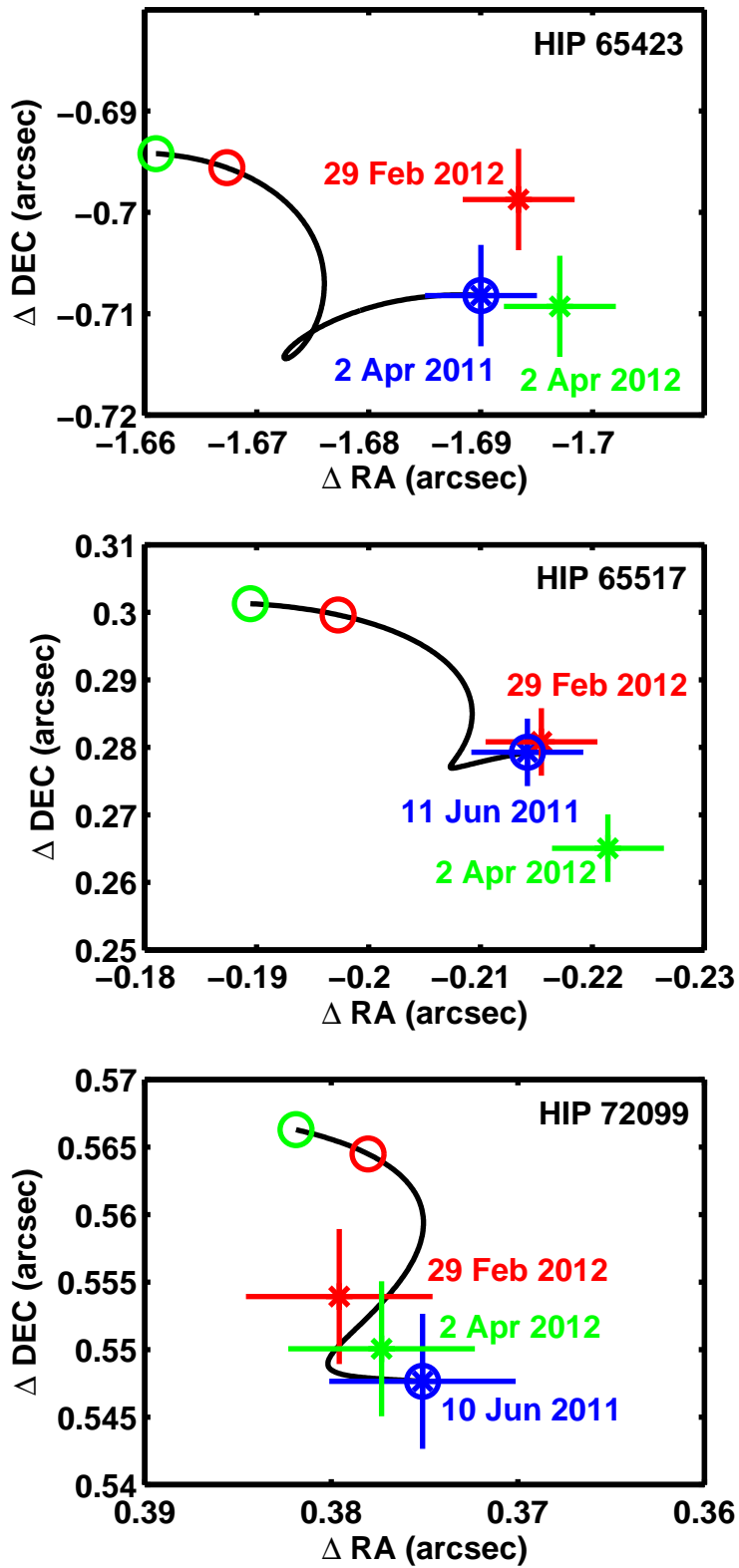}
\caption{Astrometry for each system. Symbols with error bars are the measured positions. The solid line shows the predicted motion for a static background star, with circles marking the expected location at each respective date. The background hypothesis can be excluded in each case, confirming common proper motion.}
\label{f:astro}
\end{figure*}

The astrometry of each candidate relative to the expected background motion is shown in Fig. \ref{f:astro}. The scatter is consistent in the $x$- and $y$-positions for all targets, and has an average of 5~mas, which we adopt as the astrometric uncertainty. It can be seen that each of the candidate companions is well consistent with common proper motion. For HIP~65423, the significance is 9.4$\sigma$, for HIP~65517 it is 11.0$\sigma$, and for HIP~72099 it is 4.0$\sigma$. We thus consider all of them to be confirmed companions. 

\subsection{Mass estimation}
\label{s:spectra}

Our primary objective for the spectroscopic analysis is to generate mass estimates for the companions. We do this in two independent ways: First, we calculate $H$- and $K$-band brightnesses for each companion. This is easily done since we have a direct contrast spectrum between the primary and companion. Hence, we can calculate the contrast within a given band as the mean of the contrast spectrum within that band (weighted appropriately by the spectral response curve for $H$ and $K$, respectively), and then add the measured $H$- and $K$-band magnitudes for the star to get the corresponding values for the companion. Note that extinction is negligible for all three stars \citep{chen2011}. Following this procedure, and converting to absolute magnitudes using Hipparcos \citep{perryman1997,vanleeuwen2007} parallaxes, gives $M_{\rm H} = 7.46$~mag and $M_{\rm K} = 7.04$~mag for HIP~65423~B, $M_{\rm H} = 7.63$~mag and $M_{\rm K} = 7.32$~mag for HIP~65517~B, and $M_{\rm H} = 6.90$~mag and $M_{\rm K} = 6.58$~mag for HIP~72099~B. Using the BCAH evolutionary models \citep{baraffe1998} for an age of 10~Myr \citep{song2012} gives a mass of $\sim$65~$M_{\rm jup}$ consistently in both $H$ and $K$ for HIP~65423~B. For HIP 65517~B, the mass is $\sim$55$M_{\rm jup}$ and for HIP~72099~B the mass is $\sim$95$M_{\rm jup}$ consistently in both filters.

\begin{figure*}[p]
\centering
\includegraphics[width=10cm]{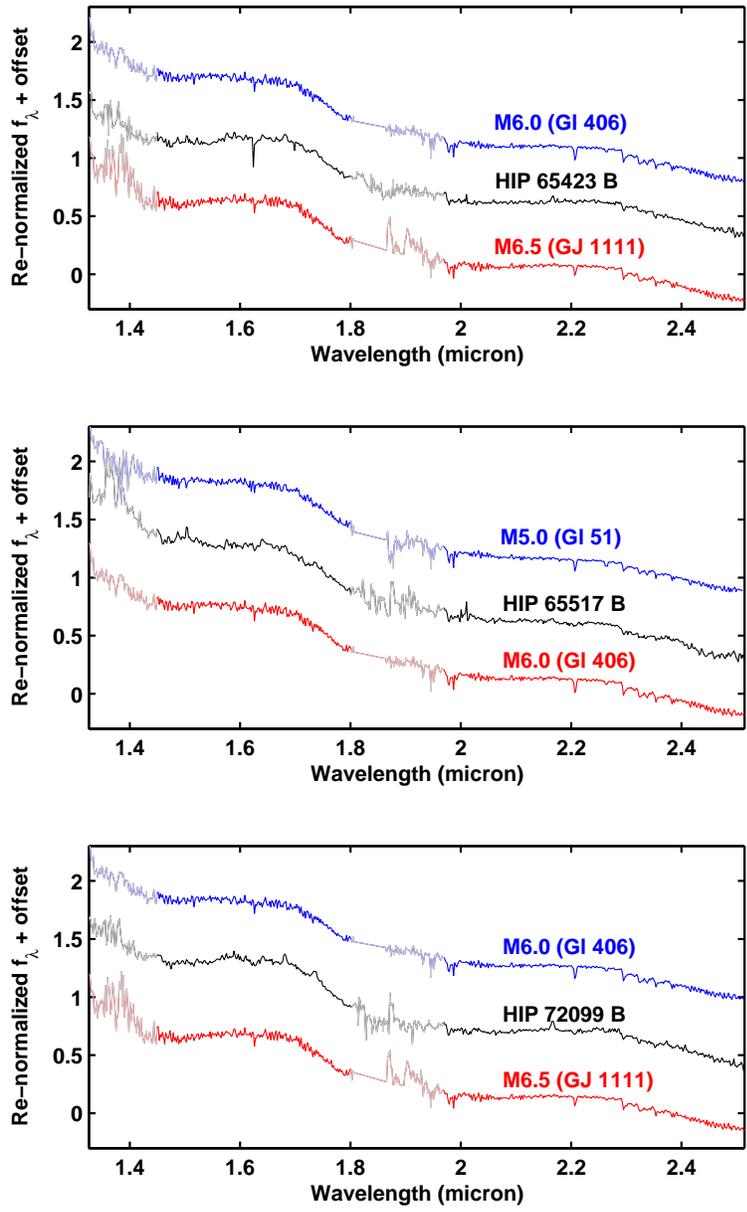}
\caption{Spectroscopy for HIP~65423~B, HIP~65517~B, and HIP~72099~B. The target spectrum in black is wedged between the two best fitting spectral templates from \citet{cushing2005} and \citet{rayner2009}.}
\label{f:spec}
\end{figure*}

For the second method, we compare the spectral continuum with library standard stars in order to determine a spectral type (SpT) and temperature. The standard stars are taken from the IRTF spectral library for M-type stars \citep{cushing2005,rayner2009}. From visual inspection, it is clear that the companions are in the SpT range of $\sim$M6, so for a formal test we use the standard stars Gl~268AB (M4.5V), Gl~51 (M5.0V), GJ~1111 (M6.0V), Gl~406 (M6.5V) and Gl~644C (M7.0V) to encapsulate the feasible physical range (there is no M5.5V standard in the library). These standard stars are probably older than our detected brown dwarfs, hence the surface gravities differ modestly, but the two classes of objects are observationally much closer to each other than to giants or sub-giants \citep[e.g.][]{slesnick2004}, and we have verified that such standard stars provide a visually worse match to our companions than the luminosity class V objects. We fit each standard star to each target spectrum and test the quality of the fit using a squared residual (effectively $\chi^2$) metric. We fit the $H$- and $K$-band ranges separately (1.49--1.83~$\mu$m and 1.97--2.33~$\mu$m, respectively), and re-normalize the target and standard to have the same mean flux value in the fitted range. In all cases (for each target and each wavelength range), the squared residuals as function of spectral type appear to be continuous functions with well-defined minima, hence we fit a second-order polynomial and adopt the spectral type at the minimum of the fitted function as the best fit. For HIP~65423~B, this gives M6.03 in $H$ and M6.31 in $K$, for HIP~65517~B, it gives M5.30 in $H$ and M5.45 in $K$, and for HIP~72099~B it gives M5.85 in $H$ and M6.50 in $K$, all nicely consistent and in accordance with visual inspection. For translating spectral types into temperatures, we use two different empirical relations \citep{golimowski2004,slesnick2004} to represent the uncertainty in this translation. The difference between the relations is $\sim$100~K at these spectral types. For some of the cases, we have to slightly extrapolate the \citet{golimowski2004} relation, since it is calibrated for $\geq$M6.0. As best-fit values we consider the mean of the spectral types derived from $H$ and $K$, and the mean of the resulting temperatures from the two relations. This gives 2770~K for both HIP~65423~B and HIP~72099~B, and 2960~K for HIP~65517~B. Using the BCAH model, we thus find a mass of $\sim$45$M_{\rm jup}$ for HIP~65423~B, $\sim$70$M_{\rm jup}$ for HIP~65517~B, and $\sim$45$M_{\rm jup}$ for HIP~72099~B. The spectra are shown in Fig. \ref{f:spec}.

Each of the methods used above have their respective uncertainties due to significant uncertainties in the underlying parameters. For the brightness-based estimate, the dominant uncertainty is in the distance estimation -- due to the small parallax signature of these stars at their $>$100~pc distances, the error is $\sim$10--20\%, leading to 0.3--0.5~mag uncertainties in the distance modulus. Our final upper and lower limits on companion mass from this method also include photometric uncertainties, as well as the uncertainty in age, where we have adopted an age range of 8--12~Myr, given that the age of Sco-Cen has been determined as being between that of TW~Hya and $\beta$~Pic \citep{song2012}. The values are provided in table \ref{t:param}.

The temperature-based estimate has the advantage of being independent of distance, but has its own uncertainties. For our error limits quoted in table \ref{t:param}, we have included the scatter in spectral types derived from the different bands, the scatter in spectral type/temperature relation, and the uncertainty in age. One possible issue that may be relevant to keep in mind for the temperature-based estimation is the fact that in the eclipsing brown dwarf binary 2M0535-05 \citep{stassun2006}, the primary (more massive) component is colder than the secondary one. The reason for this effect is still under discussion, where arguments for and against physical mechanisms such as chromospheric activity have been presented \citep[][]{stassun2012,mohanty2012}.  Our detected brown dwarfs are older than the $\sim$1~Myr 2M0535-05 system, so if the effect is age-related (e.g. due to activity or accretion), we may expect that the effect should be smaller in our case. Whatever the mechanism, the change in temperature is compensated for by an inflated radius in 2M0535-05~A, which keeps the luminosity largely constant, hence our magnitude-based estimates should be largely insensitive to this effect. In any case, the estimated masses for all three companions are consistent between the two methods, within the error bars.

\begin{table}[p]
\caption{Observed and derived parameters for the discovered systems.}
\label{t:param}

\footnotesize{
\begin{tabular}{lcccccc}
\hline
\hline
Property & HIP~65423~A & HIP~65423~B & HIP~65517~A & HIP~65517~B & HIP~72099~A & HIP~72099~B \\
\hline
SpT & G3 & M6 & G2 & M5.5 & F6 & M6\\
$T_{\rm eff}$\tablenotemark{a} (K) & 5830 & $2770^{+70}_{-50}$ & 5860 & $2960^{+150}_{-120}$ & 6360 & $2770^{+120}_{-80}$ \\
Dist. (pc) & 124.4$\pm$26.6 & ... & 110.6$\pm$18.6 & ... & 157.7$\pm$40.6 & ... \\
App. $H$ (mag)\tablenotemark{b} & 8.18$\pm$0.06 & 12.93$\pm$0.06 & 8.18$\pm$0.03 & 12.85$\pm$0.03 & 8.43$\pm$0.02 & 12.89$\pm$0.02 \\
App. $K$ (mag)\tablenotemark{b} & 8.08$\pm$0.02 & 12.51$\pm$0.02 & 8.08$\pm$0.03 & 12.54$\pm$0.03 & 8.40$\pm$0.03 & 12.57$\pm$0.03 \\
Mass\tablenotemark{c} & 1.3~$M_{\sun}$ & $65^{+35}_{-30}/45^{+10}_{-10}$~$M_{\rm jup}$ & 1.3~$M_{\sun}$ & $55^{+30}_{-20}/70^{+35}_{-15}$ ~$M_{\rm jup}$ & 1.4~$M_{\sun}$ & $95^{+65}_{-40}/45^{+15}_{-15}$~$M_{\rm jup}$ \\
Ang. sep. (\arcsec) & ... & 1.835$\pm$0.005 & ... & 0.350$\pm$0.005 & ... & 0.667$\pm$0.005 \\
PA ($^{\rm o}$) & ... & 247.4$\pm$0.2 & ... & 321.7$\pm$1.4 & ... & 34.4$\pm$0.4 \\
Proj. sep. (AU) & ... & 228$\pm$49 & ... & 39$\pm$7 & ... & 107$\pm$27 \\
\hline
\end{tabular}
}
\tablenotetext{a}{Inferred from spectral types using \citet{sherry2004} for the stars and \citet{slesnick2004} for the companions.}
\tablenotetext{b}{Stellar magnitudes from 2MASS \citep{skrutskie2006}.}
\tablenotetext{c}{Stellar masses inferred using \citet{dantona1997}, brown dwarf masses using \citet{baraffe1998}. For the brown dwarfs, the first value is based on intrinsic brightness and the second on temperature.}
\end{table}

\section{Discussion}

There is a well-known deficit of companions to stars with mass ratios in the range of $\sim$1--10\% at small semi-major axes (a few AU or less); the so-called brown dwarf desert \citep[e.g.][]{grether2006}. This desert is a natural consequence of the difficulty for stellar systems to form at small mass ratios \citep[e.g.][]{bate2009}, and the simultaneous difficulty for planetary systems to form at large mass ratios \citep[e.g.][]{mordasini2009}. It is still unclear to which extent the brown dwarf desert extends also to large semi-major axes (tens or hundreds of AU). While there are several indications that brown dwarfs are rare at these separations \citep[e.g.][]{kasper2007,nielsen2010,janson2012}, some evidently do exist. Our detected brown dwarfs have best-fit mass ratios of 4--5\%, right in the middle of the desert. Other known systems with similar mass ratios and semi-major axes include e.g. GQ~Lup \citep{neuhauser2005,janson2006,mcelwain2007} and GJ~758 \citep{thalmann2009,currie2010,janson2011} at 3--4\%. While small compared to most binary systems, these are nonetheless an order of magnitude higher than the mass ratios of directly imaged planetary systems, such as HR~8799 \citep{marois2008,marois2010} or $\beta$~Pic \citep{lagrange2009,lagrange2010}. Another interesting comparison is provided by the wide systems with companions of lower mass than our brown dwarfs that have been recently reported in the same young region: 1RXS~J1609 \citep{lafreniere2008,lafreniere2010}, HIP~78530 \citep{lafreniere2011} and GSC~0621 \citep{ireland2011}, with mass ratios of 1--2\%. Such low mass ratios by themselves may imply that these systems could have formed in a planet-like manner, perhaps through formation at a small semi-major axis and subsequent scattering outward through gravitational interactions with other companions in the system \citep[e.g.]{veras2009}. However, the fact that systems with intermediate mass ratios exist, such as those discovered here, at some non-negligible rate (perhaps $\sim$3\% if we naively divide the objects confirmed so far by the number surveyed) complicates such an analysis. Although mass ratio remains highly useful to distinguish between formation scenarios, it is not as clear-cut for wide companions as it is closer to the star. Capture of free-floating objects in young stellar cluster environments \citep{parker2012,perets2012} is a mechanism that could potentially explain an extensive range of mass ratios at wide separation. A reservoir of such objects to be captured evidently exists, down to masses below the deuterium burning limit, though possibly not in sufficient numbers \citep[e.g.][]{pena2012,scholz2012}.

\acknowledgements
We thank the staff at Gemini and ESO for their help. Support for this work was provided by NASA through Hubble Fellowship grant HF-51290.01 awarded by the Space Telescope Science Institute, which is operated by the Association of Universities for Research in Astronomy, Inc., for NASA, under contract NAS 5-26555. This study made use of CDS and SAO/NASA ADS services.

\clearpage

\end{document}